\documentclass[aps,pre,amsmath,amssymb,amsfonts,superscriptaddress,nofootinbib,a4paper]{revtex4}
\usepackage{graphicx}
\usepackage{dcolumn}
\usepackage{bm}

\def\<{\langle}
\def\>{\rangle}
\def\(({\left(}
\def\)){\right)}
\def\[[{\left[}
\def\]]{\right]}

\begin{document}

\title{Monte Carlo algorithms are very effective in finding\\
the largest independent set in sparse random graphs}

\author{Maria Chiara Angelini}
\affiliation{\footnotesize Dipartimento di Fisica, Universit\`{a} ``La Sapienza'', P.le A. Moro 5, 00185, Rome, Italy}
\author{Federico Ricci-Tersenghi}
\affiliation{\footnotesize Dipartimento di Fisica, Universit\`{a} ``La Sapienza'', P.le A. Moro 5, 00185, Rome, Italy}
\affiliation{\footnotesize INFN, Sezione di Roma1, and CNR--Nanotec, Rome unit, P.le A. Moro 5, 00185, Rome, Italy}

\date{\today}

\begin{abstract}
The effectiveness of stochastic algorithms based on Monte Carlo dynamics in solving hard optimization problems is mostly unknown. 
Beyond the basic statement that at a dynamical phase transition the ergodicity breaks and a Monte Carlo dynamics cannot sample correctly the probability distribution in times linear 
in the system size, there are almost no predictions nor intuitions on the behavior of this class of stochastic dynamics.
The situation is particularly intricate because, when using a Monte Carlo based algorithm as an optimization algorithm, one is usually interested in the out of equilibrium behavior which is very hard to analyse.
Here we focus on the use of Parallel Tempering in the search for the largest independent set in a sparse random graph, showing that it can find solutions well beyond the dynamical threshold.
Comparison with state-of-the-art message passing algorithms reveals that parallel tempering is definitely the algorithm performing best, although a theory explaining its behavior is still lacking.
\end{abstract}

\maketitle

\section{Introduction}

Discrete optimization problems defined on graphs are widespread among many scientific disciplines and commonly found in real-world applications. Depending on the properties of the underlying graph, these optimization problems may become so hard to solve that all known algorithms find only very suboptimal solutions, while the optimal ones remain unreachable to algorithms running in polynomial time.

A common benchmark to test the effectiveness of search algorithms is represented by optimization problems defined on random graphs (typical case analysis). In this case, the hardness of the optimization problem can be usually controlled by varying continuously a model parameter (e.g. the random graph mean degree or the solution size), and different algorithms can be quantitatively compared on the basis of how close to optimality they can go.

Unfortunately, in optimization problems that, in the worst case analysis, are NP-hard and also hard to approximate, a large algorithmic gap is often present in the typical case analysis, i.e.\ all known algorithms stop working at an algorithmic threshold which is bounded far away from the optimal (information theoretical) threshold. Computing the ultimate algorithmic threshold in these hard problems and understanding whether and why such an algorithmic threshold remains below the optimal one are fundamental open questions. The present work takes a step towards the answer of these questions, by studying the problem of finding a large Independent Set (IS) in a Random Regular Graph (RRG).

Given a graph $G=(V,E)$, an IS is a subset of vertices $S\subset V$ such that no vertices in $S$ are adjacent, that is $(ij)\notin E,\,\forall\,i,j\in S$.
Finding the largest IS in a graph is a fundamental problem (NP-hard in the worst case), tightly related to minimum vertex cover and maximum clique \cite{bollobas1998random}.
In physics, the problem is known under the name of hard-core model \cite{hartmann2006phase}, because vertices in $S$ can be seen as particles that have a hard-core interaction and cannot be adjacent. The largest IS thus corresponds to the densest packing configuration in the hard-core model.

We call $\rho$ the relative size of the IS, that is $|S|=\rho |V| = \rho N$.
On RRG of constant degree $d$ it has been proved that, in the large $N$ limit, IS with $\rho<\rho_{max} \sim 2 \log d / d$ do exist with high probability for $d$ large enough \cite{bollobas1976cliques,frieze1990independence}. However algorithms running in polynomial time cannot find IS with $\rho > \rho_{alg} \sim \log d /d$ for $d$ large enough \cite{coja2015independent}. And actually this algorithmic threshold $\rho_{alg}$ can be achieved with very simple algorithms \cite{grimmett1975colouring}.
The algorithmic gap, that is the strict inequality $\rho_{alg}<\rho_{max}$, has been proven for a class of local algorithms in the large $d$ limit \cite{gamarnik2014limits}. In this case, the origin of the algorithmic failure is due to the ergodicity breaking taking place at $\rho_{alg}$: this is a common phenomenon in optimization problems \cite{mezard2005clustering,achlioptas2011solution}, also called clustering or shattering of the solution space.

One expects the ergodicity breaking taking place at $\rho_{alg}$ to affect also other types of algorithms. In particular, the sampling of the optimal solutions through numerical methods based on Monte Carlo Markov Chain should become much slower when ergodicity is broken, due to the need of overcoming large barriers.
However if one is just interested in finding a single optimal or very close to optimal solution maybe Monte Carlo methods may work better than expected. This is a question never investigated in detail (to the best of our knowledge) and its answer is one of the main motivations for the present work.

We are going to analyze the performances of different algorithms, dedicating particular attention to those based on Monte Carlo Markov Chains, and we will try to relate such performances to the relevant phase transitions taking place in the space of IS in the limit of large RRG.
Indeed, studying the thermodynamics of the problem via the cavity method the authors of Ref.~\cite{barbier2013hard} showed how the space of IS changes while increasing $\rho$: for $d<16$ it undergoes a continuous phase transition from a Replica Symmetric (RS) phase to a phase described by a Full Replica Symmetry Breaking (FRSB) solution; while for $d\ge16$ the space of IS undergoes a random first-order transition (RFOT) and it can be described by a solution with one step of Replica Symmetry Breaking (1RSB).

Let us briefly review the important phase transitions in the RFOT case, each one corresponding to a drastic change in the structure of the set of ISs.
At small densities $\rho$, the ISs form a single large cluster (two ISs are considered adjacent if they differ in $o(N)$ vertices) and can be well described by an RS solution that assumes the existence of a single state.
Increasing the density, one first finds a dynamical threshold $\rho_d$ above which the space of ISs is divided into an exponential number in $N$ of distinct clusters. This is the ergodicity breaking phase transition that affects local search algorithms and Monte Carlo methods for sampling.
At the condensation threshold $\rho_c>\rho_d$ the number of clusters becomes sub-exponential, and beyond the maximum density $\rho_{max}$ there are no more ISs. This last threshold is the equivalent to the sat/unsat threshold in constraint satisfaction problems (CSP).

Beside the above thermodynamic transitions, another property has been conjectured to be important for understanding the origin of the algorithmic complexity in CSP: the concept of frozen clusters \cite{achlioptas2006solution,zdeborova2007phase,achlioptas2011solution}.
A cluster of solutions is said to be frozen if it contains frozen variables that take the same value in all the solutions of that cluster.
The rigidity threshold $\rho_r$ is defined such that for $\rho>\rho_r$ typical clusters are frozen, while above the freezing transition $\rho_f$ all clusters are frozen.
In CSP many smart algorithms can find solutions in the clustered phase, but even the most performing ones do not find frozen solutions \cite{marino2016backtracking}. For this reason, the freezing threshold is conjectured to be the ultimate algorithmic threshold.
Unfortunately, its analytic computation is a very difficult task, which has been achieved only in random hypergraph bi-coloring at present \cite{braunstein2016large}.

We will analyze different kinds of algorithms running in polynomial times. We avoid using algorithms that are known to find the largest IS in time typically growing exponentially in the graph size since these are impractical.
Three main classes of polynomial algorithms will be considered: greedy algorithms, Monte Carlo methods and message passing algorithms.
Greedy algorithms are very popular \cite{feo1994greedy,feo1995greedy,halldorsson1997greed}, because extremely fast and often provide a reasonably large IS.

We will mainly focus on Monte Carlo based algorithms that have been much less studied. Indeed the common belief is that a slow enough Simulated Annealing (SA) is able to reach densities not larger than the bottom of the equilibrium states at $\rho_d$ \cite{zdeborova2010generalization}.
Above $\rho_d$ ergodicity is broken and Monte Carlo methods should not be able to sample correctly the equilibrium properties of the model. However, it could always be possible that there are states accessible to the out of equilibrium dynamics that terminate at densities $\rho>\rho_d$ and thus an out-of-equilibrium process can find very large IS with $\rho>\rho_d$.

Recently it has been proposed to enhance the weight of deep, large states in an efficient way coupling some replicas of the system, for example in an SA algorithm \cite{baldassi2016unreasonable}. 
The Replicated SA (RSA) has been seen to enhance the performances of learning in some models of neural networks.
Here we apply RSA to the problem of finding the largest IS problem, discovering indeed that this algorithm is able to find solutions when the standard SA is not able to, well beyond $\rho_d$.
However, this seems to be true only if the transition is strongly discontinuous (RFOT). In case the transition is weakly discontinuous (or continuous) RSA and SA show similar performances.

Finally, we will analyze the behavior of Parallel Tempering (PT). Although PT has been invented to sample {\it at equilibrium} the very rough energy landscape of disordered systems and posterior distributions \cite{hukushima1996exchange,earl2005parallel}, it can be used in the {\it out-of-equilibrium} regime to try to reach some of the lowest energy configurations \cite{moreno2003finding}.
Recently the PT has been applied to the planted IS problem, allowing to find the planted configuration in the supposedly hard regime (i.e.\ when the planted IS is very small) in a time that seems to scale polynomially with the system size \cite{angelini2018parallel}.
In the random case, we show here that PT is able to find solutions above the algorithmic threshold of the SA and of all the other analyzed algorithms, included the
Belief-Propagation with Reinforcement, that is usually the best-performing message passing algorithm in other optimization problems, able to go beyond the rigidity transition \cite{dallasta2008entropy}.
We will measure the scaling of the convergence time for PT, showing that indeed it stays polynomial for $\rho>\rho_d$.

\section{Problem definition and description of analyzed algorithms}

In this section, we report the details of the problem, and of the algorithms whose performances are analyzed in the rest of the paper.

The optimization problem we try to solve is to find the largest IS in a given RRG.
Being $K$ the size of an IS, we call $\rho=K/N$ its density.
Finding the largest IS problem is clearly a zero-temperature problem since it imposes strong constraints on any pair of nearest neighbor vertices not to be in the IS.
As usual in a statistical mechanics approach, we can add a temperature parameter $T=1/\beta$ and relax the strong constraints into soft ones.
The probability measure can be written as
\begin{equation}
P(\underline{n}) \propto \exp \bigg[ \mu \sum_{i=1}^N n_i + \beta \sum_{(ij)\in E} n_i n_j\bigg]
\label{eq:measure}
\end{equation}
where $n_i\in\{0,1\}$. In the $T\to 0$ limit, vertices with $n_i=1$ form the IS, and the largest IS can in principle be achieved by sending $\mu\to\infty$ afterwards.

In practice, we are going to approach such a limit ($T\to 0$ and $\mu\to\infty$) in two different ways.
In the first way, we fix the IS size $K$, such that the first term in the measure in Eq.~(\ref{eq:measure}) is constant and can be ignored, and we study the problem in temperature.
In the second way we fix $T=0$, making constraints hard, that is we rewrite the measure as follows
\begin{equation}
P(\underline{n}) \propto \exp \bigg[ \mu \sum_{i=1}^N n_i\bigg] \prod_{(ij)\in E} (1-n_i n_j)
\label{eq:measureT0}
\end{equation}
and we study the problem increasing $\mu$.

We will use many different algorithms, described in the following list.
Each algorithm will show its own algorithmic threshold $\rho_{alg}$ above which that algorithm is not able to find IS.

\begin{itemize}
\item \textbf{Greedy algorithms} (GA)
 
Greedy algorithms are linear time algorithms where variables are set just once during the process of finding an IS. They differ according to the rule which is used to select the next vertex to include in the growing IS. Schematically they work as follows:
 \begin{itemize}
 \item start with all $n_i=0$;
 \item at each step choose a vertex $v$ from the graph and add it to the IS, i.e.\ set $n_v=1$;
 \item the vertex is chosen uniformly at random in the `random vertex' version (RV GA) and such as to have the smallest degree in the `minimum degree' version (MD GA);
 \item all the neighbors of the chosen vertex are removed from the graph.
\end{itemize}

The random vertex version has been designed by Karp and Sipser \cite{Karp1981Maximum} and produces with high probability an IS of size $N\log(d+1)/d$ both at finite and large $d$. 
The minimum degree version has been introduced in Ref.~\cite{wormald1995differential} and gives better results, at least for finite $d$, while it has the same scaling at large $d$ values.

The computational time of the greedy algorithm scales as $O(dN)$, that is linear in the graph size.

\item \textbf{Monte Carlo in temperature} ($\beta$MC)
 
We fix the size $K$ of the IS we would like to find and the temperature $T=1/\beta$ to be used in the Monte Carlo algorithm.
The algorithm will sample configurations with exactly $K$ variables set to $n=1$, that is $\sum_i n_i=K$; 
each of these configurations can be equivalently described in terms of the subset of vertices containing a particle $\mathcal{I} \equiv \{i\in V: n_i=1\}$.
To each configuration we associate the energy $E(\underline{n}) = \sum_{(ij)\in E} n_i n_j$ counting how many pairs of nearest neighbours are filled ($n=1$). 
A configuration of zero energy is an IS of size $K$.

We start by choosing $\mathcal{I}$ as a random subset of $K$ vertices of $V$.
At each step of the algorithm we propose to move a randomly chosen particle to a randomly chosen empty vertex; 
the particle and the empty vertex do not need to be nearest neighbors, so the algorithm is not standard diffusion.
Calling $\underline{n}$ the current configuration and $\underline{n}'$ the proposed configuration, we follow standard Metropolis rule for accepting the proposed configuration, 
that is we accept the change with probability 1 if $E(\underline{n}') \le E(\underline{n})$, and with probability
$\exp[\beta(E(\underline{n}')-E(\underline{n}))]$ otherwise.
As done conventionally, we define a Monte Carlo Sweeps (MCS) the attempt to move a randomly chosen particle, repeated $K$ times.
We stop the algorithm when a configuration $\underline{n}_{IS}$ with $E(\underline{n}_{IS})=0$ is found, that corresponds to an IS.

\item \textbf{Parallel Tempering in temperature} ($\beta$PT)

We consider $N_\beta$ replicas, each one with exactly $K$ variables set to $n=1$ as in the $\beta$MC method discussed above. Each replica undergoes a standard Metropolis evolution at inverse temperature
$\beta_i=\beta_{max}-i\cdot \Delta\beta$, $i\in[0,N_\beta-1]$. Every 5 steps of $\beta$MC a temperature swapping step is attempted for each pair of configurations at nearby temperatures $\beta_i$ and $\beta_{i+1}$; the temperature swap is accepted with probability 
\begin{equation}
p=\text{min}\((1,e^{\((\beta_i-\beta_{i+1}\))\((E_i-E_{i+1}\))}\)),
\label{eq:flipping}
\end{equation}
where $E_i$ is the current value of the energy of the $i$-th replica.
The algorithm is stopped when a replica (usually the one with the lowest temperature) reaches a zero energy configuration.

\item \textbf{Simulated Annealing in chemical potential} ($\mu$SA)
 
Working directly at zero temperature, i.e. sampling the measure in Eq.~(\ref{eq:measureT0}), we run a Simulated Annealing scheme in the following way.
We start from the empty configuration $n_i=0\;\forall i$ that certainly satisfy all the constraints and from a null chemical potential $\mu=0$.
At each step of the SA algorithm we increase the chemical potential by $\Delta\mu$ and we do a Monte Carlo sweep, that corresponds to the attempt to update each of the $N$ variables $n_i$ following the usual Metropolis rule: in practice if $n_i=0$, we set $n_i=1$ only if all the nearest neighbors are empty, and if $n_i=1$ we set $n_i=0$ with probability $\exp(-\mu)$. 
We stop the SA algorithm at a value $\mu_{max}$ where we observe the IS density $\rho=\sum_i n_i /N$ not increasing any more on any reasonable timescale.

The algorithm, at fixed parameter $\Delta\mu$, is linear in the size $N$.
 
 \item \textbf{Replicated Simulated Annealing in chemical potential} ($\mu$RSA)
 
In Ref. \cite{baldassi2016unreasonable} a replicated version of the SA is proposed to sample with higher probability states with larger entropy.

To define the Replicated SA, we introduce $R$ replicas of the variables on the same RRG, and a coupling between the different replicas according to the following measure:
\begin{equation}
P(\underline{n}^1,\ldots,\underline{n}^R) \propto \exp \bigg[ \mu \sum_{a=1}^R \sum_{i=1}^N n_i^a + \gamma \sum_{a< b} \sum_{i=1}^N n_i^a n_i^b\bigg] \prod_{a=1}^R \prod_{(ij)\in E} (1-n_i^a n_j^a)
\label{eq:repH_HC}
\end{equation}
We then run the SA algorithm on this replicated system, fixing the value of $\gamma$ and incrementing the value of $\mu$ as in the $\mu$SA. At variance to numerical experiments in Ref.~\cite{baldassi2016unreasonable}, where $\gamma$ is incremented during the annealing, we prefer to keep $\gamma$ fixed as we have seen that varying $\gamma$ does not improve the final result.
 
 \item \textbf{Parallel Tempering in chemical potential} ($\mu$PT)

We consider $N_\mu$ replicas of the system, each replica being at a different chemical potential: $\mu_i=\mu_{max}-i\cdot \Delta\mu$, $i\in[0,N_\mu-1]$. 
For each replica, we run 5 Metropolis Monte Carlo sweeps at the corresponding chemical potential and then we try to swap configurations between close by values of the chemical potential with probability 
\begin{equation}
p=\text{min}\((1,e^{\((\mu_i-\mu_{i+1}\))\((-K_i+K_{i+1}\))}\)),
\label{eq:flippingPT}
\end{equation}
where $K_i$ is the actual number of variables set to 1 in the $i$-th replica.
We stop the simulation if a replica (usually the one of index 0) reaches the IS size $K$ we aim at.
 
 \item \textbf{Belief Propagation with Reinforcement} (BPR)
 
The Belief Propagation equations for the present problem were already derived in Ref.~\cite{barbier2013hard}:

\begin{equation}
 \pi_{i\to j}=\frac{e^{\mu}\prod_{k\in\partial i\backslash j}(1-\pi_{k\to i})}{1+e^{\mu}\prod_{k\in\partial i\backslash j}(1-\pi_{k\to i})},
\end{equation}

where $\pi_{i\to j}$ is the probability to have $n_i=1$ in a modified graph where edge $(ij)$ has been removed.

These equations for $\mu<\mu_c$ converge to a homogeneous paramagnetic fixed point (FP).
To turn the BP equations into a solver, one can add a reinforcement term, initially introduced in Ref.~\cite{braunstein2006learning}, with two parameters $\gamma$, $dt$ 
that tune respectively the strength and the speed of update of the reinforcement term. Practically the eqs. for the update of the messages becomes:

\begin{equation}
\pi^{t+1}_{i\to j}=\frac{e^{\mu}[\theta_i(t)]^{1-\gamma_t}\prod_{k\in\partial i\backslash j}(1-\pi^t_{k\to i})}{1+e^{\mu}[\theta_i(t)]^{1-\gamma_t}\prod_{k\in\partial i\backslash j}(1-\pi^t_{k\to i})},
\end{equation}
with $\theta_i(t)=\prod_{k\in\partial i}(1-\pi^{t-1}_{k\to i})$ and $\gamma_t=\gamma^{\lfloor t\ dt\rfloor}$.

The FP reached when reinforcement is present is a completely magnetized one, that is the marginal probabilities for the values of $n_i$ are such that $P[n_i=1]\in\{0,1\}$, and thus each variable is surely in the IS or surely outside of it.
Thus the FP reached by BPR does correspond to an IS.
 
\end{itemize}

The attentive reader probably notices that the above list is not including all possible Monte Carlo schemes: for example (Replicated) Simulated Annealing in temperature and simple Monte Carlo in chemical potential are missing. For this reason, we spend now a few words discussing our choice of the analyzed algorithms and explaining how the present work is organized.

The first algorithms in the list are greedy algorithms. They are clearly suboptimal and have been run just to give an idea of the IS size which is very easy to find in linear time.
This information will also be useful to set the parameters of more refined algorithms as PT.
The algorithmic thresholds for the GA, as for all the other analyzed algorithms, are reported in Table \ref{Tab:rho_max}. 

We then analyze in Sec.~\ref{Sec:betaalg} the algorithms at fixed density, $\beta$MC and $\beta$PT. In these algorithms, the size of the IS one is looking for is fixed to $K$ and what is changed is the inverse temperature parameter $\beta$, that in turn varies the number of links within the set $\mathcal{I}$ representing the putative IS. Naturally, in the limit $\beta\to\infty$, no more links inside $\mathcal{I}$ are allowed and we obtain a true IS.
We start the discussion about stochastic algorithms with the analysis of $\beta$MC because this is an adaptation to the IS problem of commonly used local search algorithms, e.g.\ WALKSAT or ASAT \cite{aurell2004WALKSAT,ardelius2006ASAT}, which have been applied with success to problems like random $K$-SAT or random graph coloring: the main difference being that $\beta$MC respects the detailed balance condition, while WALKSAT or ASAT do not.
We then study the $\beta$PT algorithm because this is the most common way to improve Monte Carlo sampling methods in glassy systems.
This algorithm seems to scale superlinearly, but still polynomially, with the problem size $N$ as shown in detail in Sec.~\ref{Sec:scalingN}.

Then in Sec.~\ref{Sec:T0alg} we move to analyse stochastic algorithms that work directly at zero temperature, $\mu$SA, $\mu$RSA and $\mu$PT, where links inside $\mathcal{I}$ are not allowed and the tuning parameter is the chemical potential $\mu$. 
We do not study the $\beta$SA because the extrapolation of the algorithmic threshold in that case is a long and difficult task \cite{Budzynski18}: one should find the threshold for any given $\mu$ and then extrapolate int the $\mu\to\infty$ limit. The extrapolation of the algorithmic threshold is instead direct for the $\mu$SA algorithm, and for this reason, we prefer to study this version of SA.
We will see that the $\mu$PT have an algorithmic threshold similar to the $\beta$PT one, thus showing that the performances of PT are rather robust.

Finally in Sec.~\ref{Sec:BPR} we compare the results obtained via the stochastic algorithms with the outcome of BPR, which is a powerful message passing algorithm, widely used to solve problems defined on random graphs.

\begin{table}[t]
\resizebox{\textwidth}{!}{%
\begin{tabular}{ |l||l|l| l | l | l | l |l|l|l|l|l|}
    \hline
  $d$  &$\rho_d$& $\rho_c$ & $\rho_{max}$ & RV GA & MD GA & $\beta$MC & $\beta$PT & $\mu$SA & $\mu$RSA & $\mu$PT & BPR \\
     \hline
     \hline
20 & 0.1830 & 0.1833 & 0.1948 & 0.1512(1) & 0.1737(1) & 0.1906(4) & 0.1943(2) & 0.1937(1) & 0.19370(5) &  0.1945(1) & 0.1933(1) \\
 \hline
100 & 0.0638 & 0.0664 & 0.0674& 0.0447(1) & 0.0572(2) & 0.0642(1) & 0.0657(1) & 0.06470(3) & 0.06479(1) & 0.0655(1) & 0.0650(1) \\
 \hline
 \end{tabular}
 }
 \caption{Relevant physical thresholds $\rho_d$, $\rho_c$ and $\rho_{max}$ reported from Ref.~\cite{barbier2013hard} and the algorithmic thresholds found in this work for many different algorithms searching for the 
 largest IS in a RRG of degree $d=20$ and $d=100$ }
 \label{Tab:rho_max}
\end{table}

We will mainly analyze the problem at $d=20$, where the transition is still near to the continuous one, and $d=100$ where the transition is distinctly 1RSB.
In Table~\ref{Tab:rho_max}, the values for the $\rho_d$, $\rho_c$ and $\rho_{max}$, together with the thresholds for the maximum density reached by the analyzed algorithms are reported.

\section{Maximum density reached by fixed-density algorithms}\label{Sec:betaalg}

\begin{figure}
\centering
\includegraphics[width=0.47\textwidth]{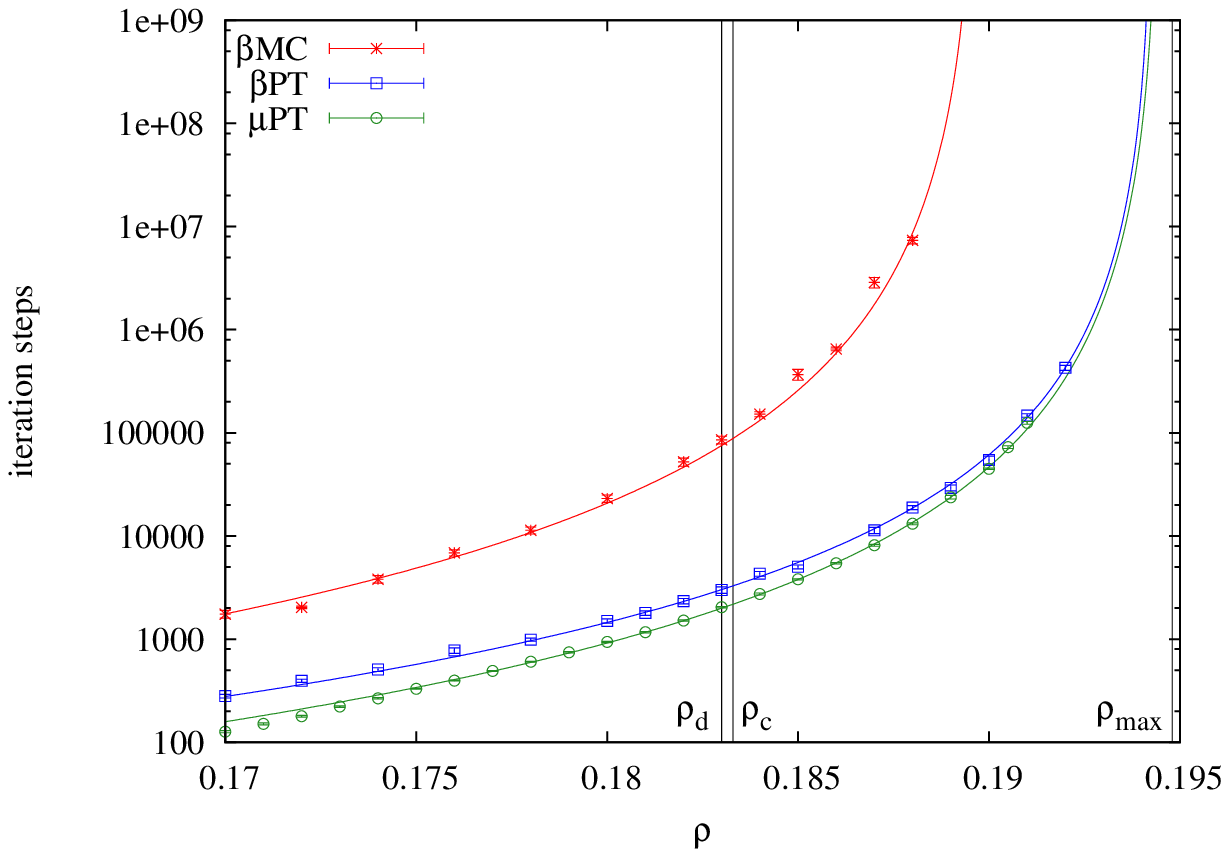}\hfill
\includegraphics[width=0.47\textwidth]{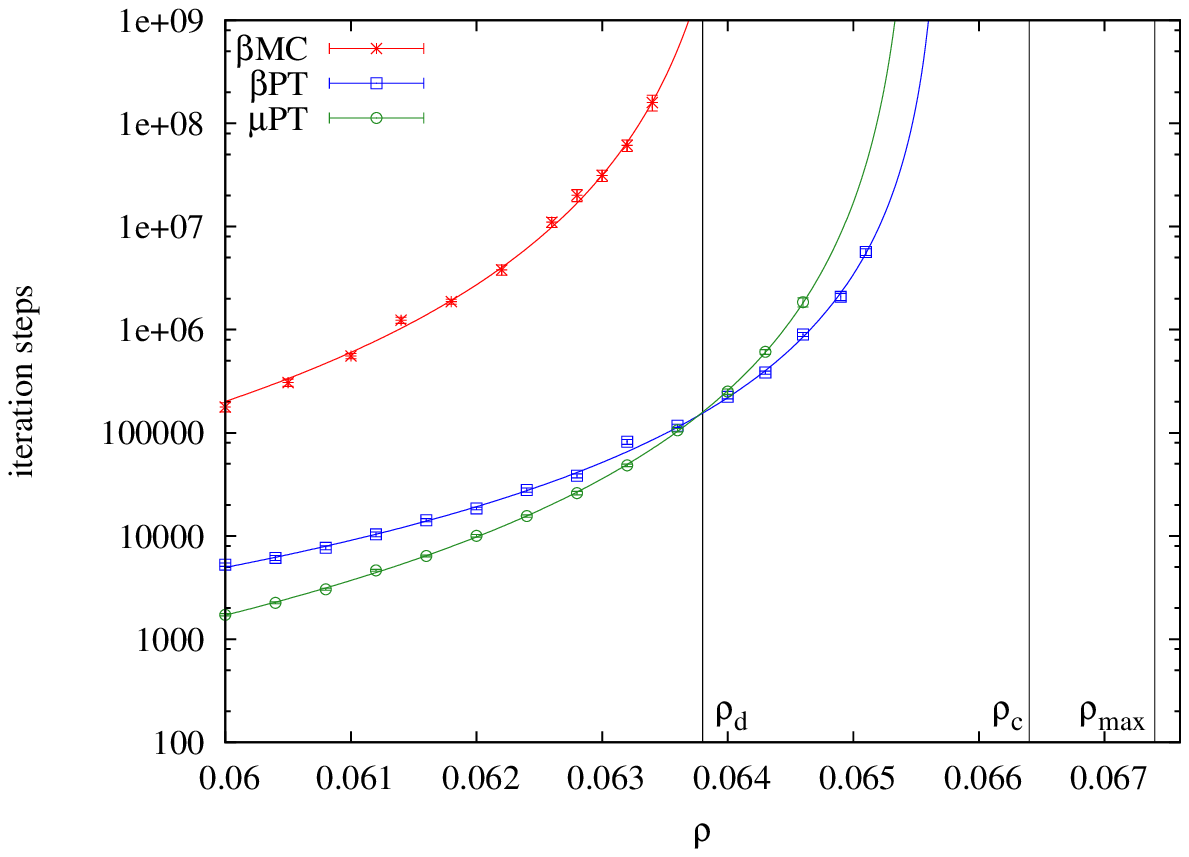}
\caption{\label{Fig:T_d20} 
Convergence time for $\beta$MC (with parameter $\beta=11$), $\beta$PT (with parameters $\beta_{max}=11$, $\Delta\beta=0.4$, $N_\beta=20$) 
and $\mu$PT algorithm (with parameters  $\mu_{max}=6$, $\Delta\mu=0.2$, $N_\mu=20$) for $N=5\cdot10^4$ and $d=20$ (left) or $d=100$ (right). The vertical lines show the theoretical thresholds
for comparison.}
\end{figure}

In this section we look at the performances of the fixed-density algorithms, namely $\beta$MC and $\beta$PT.
For these algorithms, if we measure the running time in Monte Carlo Sweeps (MCS) 
a linear dependence on $N$ is hidden in the single MCS (that takes a time proportional to $N$) 
and we can limit ourselves to measure the number of MCS needed to reach the wanted solution in order to understand the computational complexity of this class of algorithms.
In Fig.~\ref{Fig:T_d20} we show the number of MCS needed by $\beta$MC and $\beta$PT to converge to an IS of a given density $\rho$. Also, the results for $\mu$PT are shown for comparison.

For what concerns $\beta$MC, the optimal value of $\beta$ maximizing the probability of reaching an IS, i.e.\ a zero energy configuration, is likely to depend on $N$. 
Consequently, the convergence time will depend on $N$, since we expect the Monte Carlo dynamics to slow down when the temperature is decreased.
Nevertheless, we are not going to make this detailed study, because, as shown in Fig.~\ref{Fig:T_d20}, 
standard Monte Carlo run at a single temperature is easily outperformed by Parallel Tempering.

The time to find an IS of a given density $\rho$ is clearly diverging approaching the algorithmic threshold $\rho_{alg}$. 
In order to estimate the algorithmic threshold we need to perform an extrapolation. The best data interpolation is obtained via a power law divergence
\begin{equation}
\tau=\frac{C}{(\rho_{alg}-\rho)^\nu}\;,
\label{eq:tau(rho)}
\end{equation}
where $C$, $\nu$ and $\rho_{alg}$ are the fitting parameters (specific to each different algorithm).
The best fitting curves are shown with full lines in Fig.~\ref{Fig:T_d20}.
The extrapolated algorithmic threshold are reported in Table \ref{Tab:rho_max}, while the best fitting values for the $\nu$ exponent can be found in Table \ref{Tab:nu}.
Data in Fig. \ref{Fig:T_d20} are for size $N=5\cdot 10^4$, that is large enough that finite size effects are not present in the estimation of $\rho_{alg}$.
The dependence of $C$ and $\nu$ on the size will be discussed in Sec.~\ref{Sec:scalingN}.
We notice that both versions of PT (in temperature and chemical potential) have very similar algorithmic thresholds. 
This may suggest that at that density value there is some unavoidable hardness that affects both versions of PT. 
Our PT scheduling is not particularly optimized on purpose, because we believe that if an unavoidable algorithmic barrier arises at a certain density value, this should affect any version of Monte Carlo based algorithms.
The only  parameter that we decide to fix in an (almost) optimal way is $\beta_{min}$, i.e.\ the lowest value for the inverse temperature: indeed a too low $\beta_{min}$ 
requires a larger running time without any performances improvement (too many replicas at high temperature are useless), while a too large $\beta_{min}$ 
does not allow the configurations to decorrelate fast enough. We find that a very good choice for $\beta_{min}$ is the inverse temperature such 
that the actual density of the larger IS among the $K$ variables with $n=1$ is almost the maximum IS density reached by the best greedy algorithm. 
This means that the replica at $\beta_{min}$ can easily travel in the whole configurational space and this is enough for the PT algorithm to work properly.

\begin{table}
\begin{center}
\begin{tabular}{ |l||l|l| l |}
    \hline
     & $\beta$MC & $\beta$PT & $\mu$PT \\
     \hline
     
$d=20$ & 4.2(2) & 3.12(4) & 3.34(7)\\
 \hline
$d=100$ & 4.0(1) & 4.2(2) & 3.2(1) \\
 \hline
 \end{tabular}
 \end{center}
 \caption{Fitting the divergence of the convergence time shown in Fig. \ref{Fig:T_d20} via the power law $\tau=C(\rho_{alg}-\rho)^{-\nu}$, the best fitting values for $\nu$ are the ones shown in this table.}
 \label{Tab:nu}
\end{table}

\subsection{Scaling with $N$ for the $\beta$PT}\label{Sec:scalingN}

We have seen that the PT algorithm is able to find solutions in a region of $\rho$ where other algorithms fail.
The next important question to answer is how the number of PT iterations needs to be scaled with $N$ in order to find an IS of density $\rho$. The issue is particularly relevant above $\rho_d$ and approaching $\rho_{alg}$ where the convergence time diverges.
To analyze the scaling with $N$, we implement an optimized choice of the temperatures in the PT algorithm, whose derivation is in Appendix \ref{App}. 
The optimized temperatures scheduling requires a number of replicas in a range $\beta\in[0,\beta_{max}]$ that scales as $\sqrt{N}$. However, the replicas in the range $\beta\in[0,\beta_{min}]$ are useless and can be safely ignored without altering PT performances. In practice we end up with $N_\beta\sim40$ in the worst case studied ($d=100$, $N=10^5$ and $\rho=0.0646$).

 \begin{figure}
\centering
\includegraphics[width=0.5\textwidth]{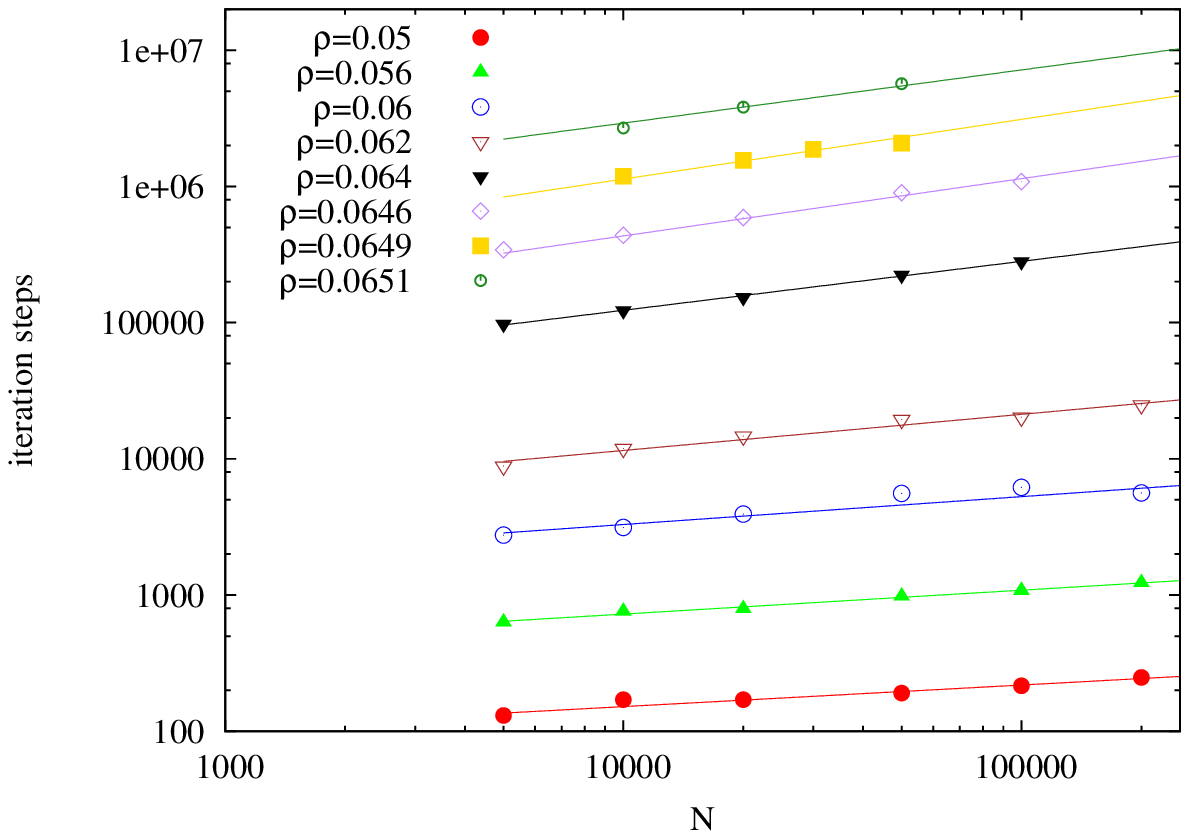}\hfill
\includegraphics[width=0.46\textwidth]{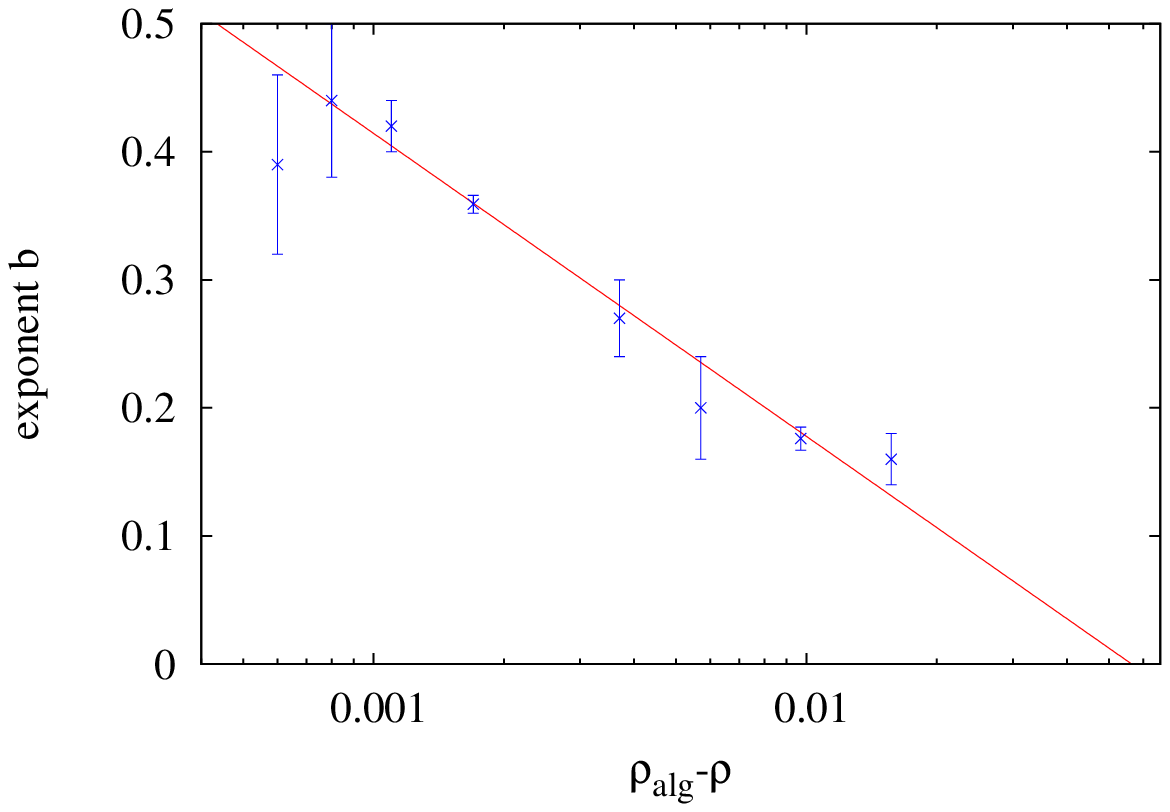}
\caption{Left: MCS to find a solution for $d=100$ for different values of $\rho$ as a function of the size $N$ of the graph for the optimized $\beta$PT. Errors are smaller than points.
The fits are of the kind $\tau(N)=a N^{b}$. Right: Dependence of the exponent $b$ on the distance from the algorithmic threshold $\rho_{alg}-\rho$. The fit is of the type 
$b=c_1+c_2\cdot\log(\rho_{alg}-\rho)$. The right border of the plot corresponds to $\rho=0$.}
\label{Fig:scalingN}
\end{figure}
 
To study the size dependence of the convergence time, we run all our $\beta$PT simulations with the temperature set defined in Eq.~(\ref{eq:betasPT}) with $r=r_\text{opt}$, between $\beta_{min}$ and $\beta_{max}$.
In Fig.~\ref{Fig:scalingN} we show for $d=100$ the results in a wide range of densities (similar behaviour is observed for $d=20$). The running times grow as a power law in $N$
\begin{equation}
\tau(N)=a(\rho)\cdot N^{b(\rho)}\,,
\label{eq:tau(N)}
\end{equation}
where the main $\rho$ dependence is in the prefactor $a(\rho)$ that diverges at $\rho_{alg}$ as in Eq.~(\ref{eq:tau(rho)}).
However there is also a slight dependence on $\rho$ in the exponent $b$. We plot $b$ as a function of $\rho$ in the right panel of Fig.~\ref{Fig:scalingN}, together with a fit of the type $b(\rho)=c_1+c_2\cdot\log(\rho_{alg}-\rho)$, that interpolates nicely the data.
We notice that this behaviour is the one to make Eqs.~(\ref{eq:tau(rho)}) and (\ref{eq:tau(N)}) compatible, since they are particular cases of the more general expression
\begin{equation}
\log(\tau(\rho,N))=\log(c)-\nu'\log(\rho_{alg}-\rho)+c_1\log(N)+c_2\log(N)\log(\rho_{alg}-\rho)\,.
\label{eq:tau(N,rho)}
\end{equation}
For a fixed value of $N$ we recover Eq.~(\ref{eq:tau(rho)}) with $C=c N^{c_1}$ and $\nu=\nu'-c_2 \log(N)$.

From the data shown in Fig.~\ref{Fig:scalingN} it is evident that the exponent $b$ is positive even in the ``easy'' region and it seems to go to zero only for $\rho\simeq0$. This means that using PT to find IS always requires a running time growing more than linearly in $N$. We think this is due to the fact that PT is a sophisticated algorithm developed to find solutions when the energy landscape is complex. For $\rho<\rho_d$, when there is just a single state, PT is thus suboptimal (maybe with a different choice of the parameters it could become a linear algorithm in this region, this kind of optimization is, however, out of our scope: we introduced PT to reach solutions in the hard region).

The time divergence as a power law approaching a given density, as in Eq.~(\ref{eq:tau(rho)}), is reminiscent of what happens in a first order phase transition, thus suggesting that at $\rho_{alg}$ an extensive barrier develops that makes impossible to reach states with $\rho>\rho_{alg}$ in polynomial time. The weak dependence on $N$, instead, suggests that some long range correlations may develop in the states in which the dynamics fall for $\rho<\rho_{alg}$ (this is discussed in the next section).

\section{Looking at the freezing}

As already mentioned, it has been conjectured for other optimization problems that the threshold for the appearance of hardness in polynomial time algorithms corresponds to the freezing threshold, that is the lowest density such that all clusters are frozen.
We want to check this conjecture in the present problem.

For practical purposes let us define a \emph{cluster} as the set of solutions (i.e.\ valid ISs) that are ``connected'' via paths where each step is the flip of just two variables.
A cluster of solutions is frozen if it contains frozen variables, that is if there is at least one variable fixed to a given value in all the configurations of the cluster. 
Above the rigidity threshold almost all the dominant clusters are frozen (but clusters with larger internal entropy might be not frozen).
The freezing threshold corresponds to the density at which each cluster of solutions is frozen.

\begin{figure}[t]
\centering
\includegraphics[width=0.6\textwidth]{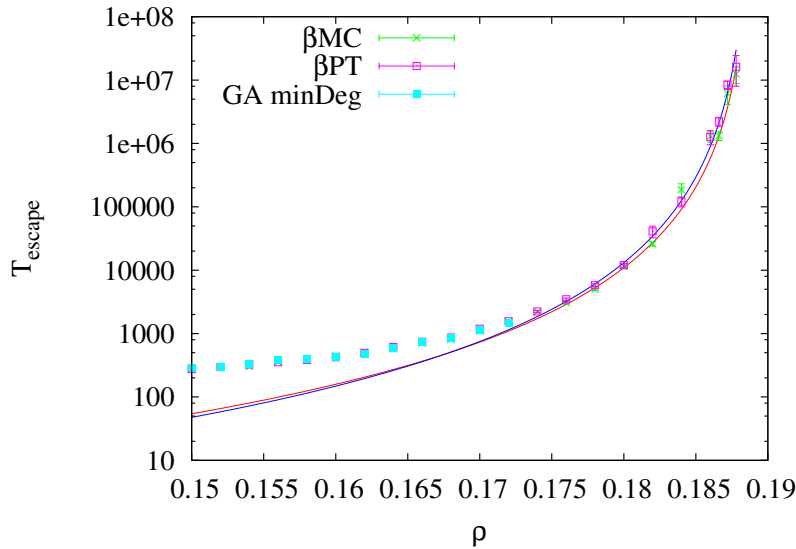}
\caption{\label{Fig:T_esc} Escape time from the solution reached by different algorithms ($d=20$, $N=5\cdot10^4$).}
\end{figure}

In this section we study the \emph{escape time}, $t_{esc}$, which is the time needed by an algorithm that moves only between solutions to go away from the initial configuration.
More precisely, we first find a solution with a given algorithm, then we apply the $\beta$MC algorithm at $\beta=\infty$ (that is a kind of diffusive dynamics at fixed zero energy and fixed size of the IS) and we measure the time needed to ``free'' each variable from its starting value, that is to find that variable in a value different from the starting one.
Looking at Fig.~\ref{Fig:T_esc}, the first important observation is that all the analyzed algorithms show the same $t_{esc}$ at a fixed density of the IS. This means that they all find the same kind of solutions (when they can find one).

\begin{figure}[t]
\centering
\includegraphics[width=0.47\textwidth]{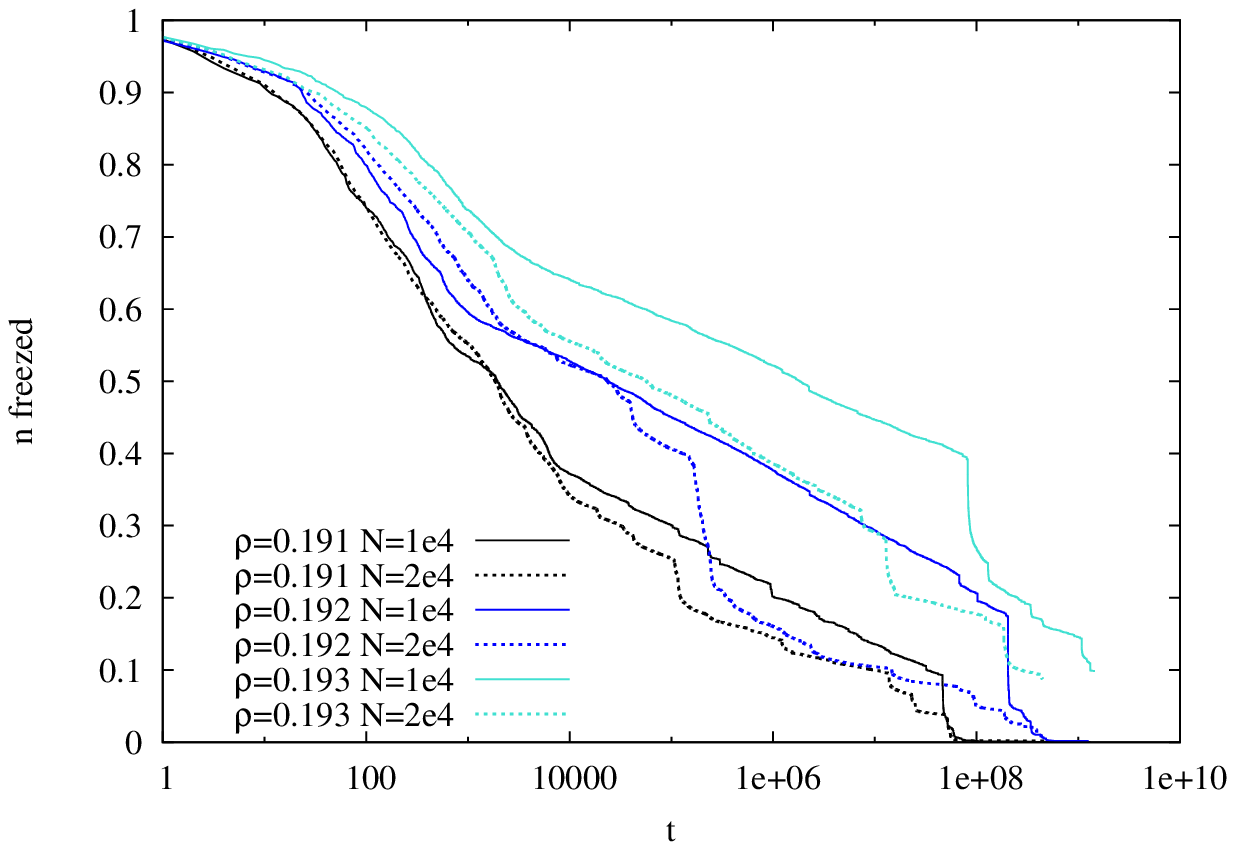}\hfill
\includegraphics[width=0.47\textwidth]{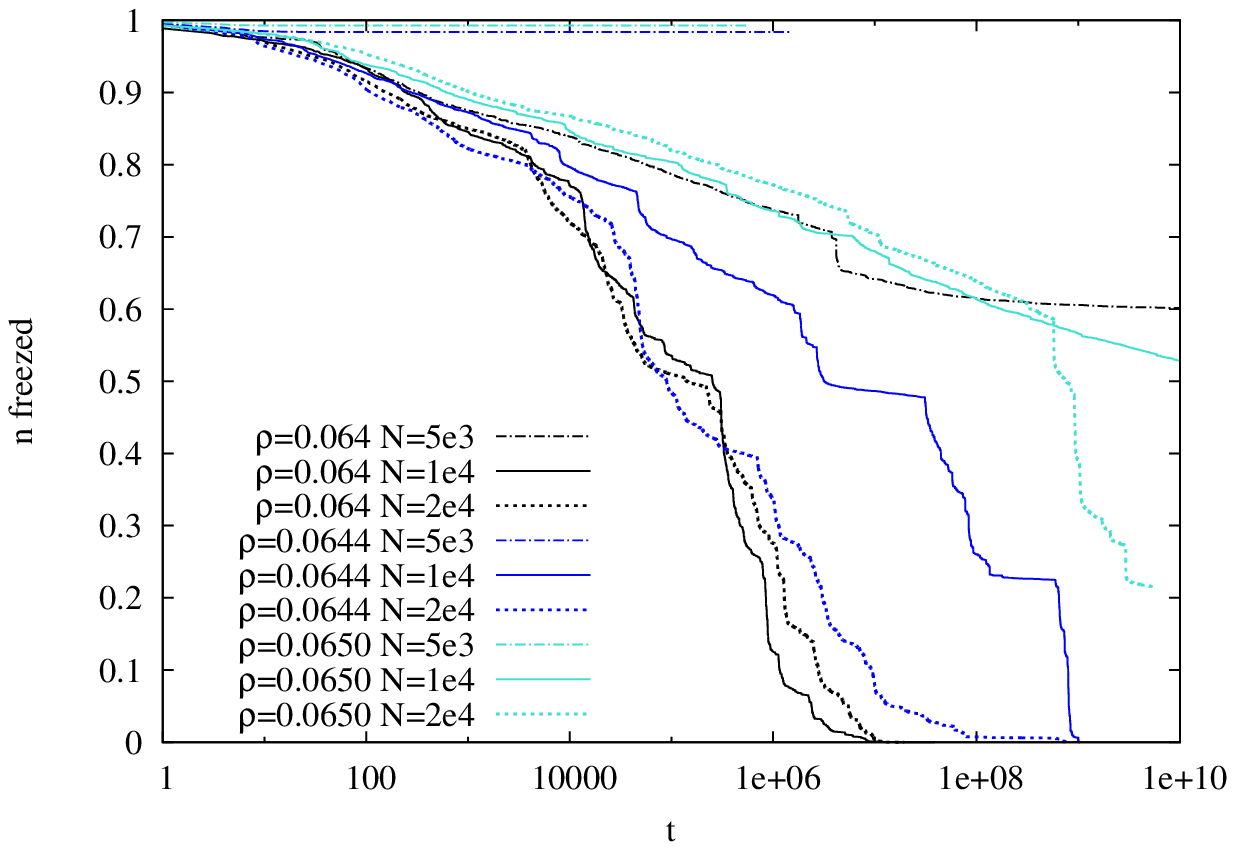}
\caption{Starting from an IS of density $\rho$ found via the $\beta$PT algorithm, we measure the fraction of variables that have not changed their value during a pure diffusive dynamics ($\beta$MC algorithm with $\beta=\infty$). Results are for $d=20$ (left), $d=100$ (right) and a single sample of the size indicated in the legend. }
\label{Fig:n_freezed_d20}
\end{figure}

The escape time diverges as a power law at a threshold density $\rho_r$ (see the fits in Fig.~\ref{Fig:T_esc}).
From the data we estimate $\rho_r(d=20)=0.1890(6)$ and $\rho_r(d=100)=0.0639(2)$.
The observation that the same threshold holds for different kind of algorithms suggests us to conjecture that $\rho_r$ does actually correspond to the rigidity threshold, that is the density where the typical clusters become frozen and the escape time from it thus diverges.

The values of $\rho_r$ are compatible with the thresholds for the $\beta$MC algorithm, while $\beta$PT and $\mu$PT can find solutions of densities greater than $\rho_r$.
At this point, it is natural to check whether the solutions found by the PT algorithms at densities larger than $\rho_r$ are frozen or not.
To answer this question we find a solution at density $\rho>\rho_r$ with the $\beta$PT algorithm, then we run the $\beta$MC algorithm at $\beta=\infty$ (the diffusive algorithm) and we look at the persistence, that is the fraction of variables that have not changed during the diffusive dynamics.

The results are shown in Fig.~\ref{Fig:n_freezed_d20} for a single sample: the fraction of frozen variables seems to decrease in an extremely slow way, mostly logarithmically in time with evident jumps (corresponding to avalanches of variables that are set free altogether).
It is worth noticing that the slowness of the diffusive dynamics around the initial solution found by PT is only due to entropic effects, given than the diffusive dynamics keeps the energy constant.

In Fig.~\ref{Fig:n_freezed_d20} we also notice some interesting finite size effects.
For the largest sizes, the diffusive dynamics eventually makes every variable unfrozen, although the escape time is some orders of magnitude larger than the time needed by PT to reach that particular solution (suggesting that PT follows a smart path that is not affected by entropic barriers!).
For smaller sizes, frozen variables persist longer and eventually we observe that the diffusive dynamics is not able to leave the cluster: the fraction of frozen variables becomes constant in time.
This is a strong evidence that the IS found by PT for small enough $N$ belong to frozen clusters (a similar phenomenon has been observed also in other models when solved for example via the Reinforcement algorithm \cite{zdeborova2009statistical}).

The above observations support the following scenario: the PT algorithm is able to find IS beyond the rigidity threshold $\rho_r$ and in this rigid phase, for small enough sizes, there is a non zero probability that PT finds a  solution in a rare frozen cluster. However, for large $N$, the solutions found by PT seems to be all unfrozen and thus we deduce that the PT algorithmic threshold is bounded above by the freezing threshold. We are strongly tempted to conjecture that the two thresholds, $\rho_{alg}$ for PT and $\rho_f$, do actually coincide, but we do not have firm arguments in support.

We have also checked that the solutions found by the PT algorithms above $\rho_d$ are not equilibrium solutions.
To do this, we find a solution at $\rho>\rho_d$ with the $\beta$PT or $\mu$PT algorithms. We then initialize BP on that solution and we check whether BP converges to a fixed point close to the solution found by PT. If it is so, this means that the PT solution lays inside one of the states (and replica symmetry holds within a state) that form the 1RSB structure that characterises the equilibrium measure for densities slightly above $\rho_d$. 
However, we find that BP does not converge (neither to the paramagnetic fixed point nor to a fixed point close to the PT solution). This lack of convergence suggests that the solution found by PT is probably inside a state that is not replica symmetric, but probably FRSB, as found in other models \cite{zdeborova2010generalization}. Indeed it is well known that states reached by the out-of-equilibrium dynamics may be FRSB even when equilibrium states are 1RSB \cite{montanari2003nature,montanari2004cooling}.

\section{Zero temperature algorithms} \label{Sec:T0alg}

In this section, we analyze a different kind of algorithms, the ones running directly at zero temperature. This means that links inside $\mathcal{I}$ are not allowed, i.e.\ the algorithm always works with a valid IS. For this class of algorithms, the varying parameter is the chemical potential $\mu$ that changes the average density of the  IS. The limit $\mu\to\infty$ should correspond to the largest possible IS.

First of all, we run $\mu$SA. It is a common belief that a slow enough SA should reach the bottom of the equilibrium states at $\rho_d$. 
The algorithmic thresholds, computed as the average over $100$ samples of the maximum density $\rho$ reached when $\mu\to\infty$ in a SA with $\Delta\mu=10^{-7}$ and $N=5\cdot 10^4$, are reported in Table \ref{Tab:rho_max}.
As one can notice, for $d=20$ the inequalities $\rho_{alg}>\rho_c>\rho_d$ hold, implying that the states that dominate the measure at $\rho_d$ can be followed deeply beyond $\rho_c$. This is compatible with the fact 
that at $d = 20$ the transition is still near to a continuous FRSB one and thus the ergodicity breaking is less pronounced. For $d=100$ instead $\rho_{alg}<\rho_c$, consistently with the fact that the transition is distinctly 1RSB and ergodicity breaking takes place in a much more marked way.

We then pass to analyze the $\mu$RSA algorithm. We take inspiration from Ref.~\cite{baldassi2016unreasonable},
where a replicated version of the SA is proposed to sample with higher probability states with larger
entropy. In the context of ISs one can identify a state in the following way: starting from a maximal IS,
that is an IS that cannot be increased any further by just adding vertices to the IS itself, and considering
this maximal IS as the ``bottom of a valley'' in a usual energy landscape, one can build a state by the set of
the ISs which are a subset of the maximal one (the construction has to be refined when one finds ISs which
are a subset of more than one maximal IS, but we do not need such a detailed description for the present
argument). According to this construction it is likely that states corresponding to the largest IS are also
those of largest entropy. So the use of an algorithm that favours states according to their entropy is likely
to be beneficial also in the search for the largest IS.

\begin{figure}[t]
\centering
\includegraphics[width=0.5\textwidth]{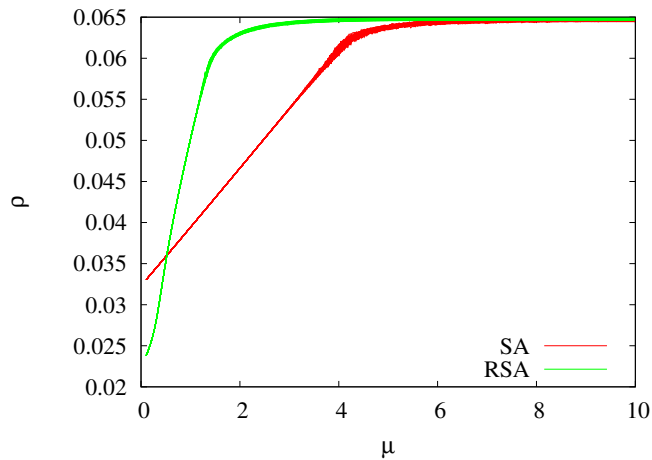}
\caption{\label{Fig:SA} Comparison between $\mu$SA and $\mu$RSA for 50 samples of size $N=5\cdot 10^4$ and $d=100$ (parameters are $\Delta\mu=10^{-7}$, $R=3$ and $\gamma=1$).}
\end{figure}

We run $\mu$RSA with parameters $R=3$, $\gamma=1$.
In Fig.~\ref{Fig:SA} its performances are compared with those of $\mu$SA in the case $d=100$ (their algorithmic thresholds can be found in Table~\ref{Tab:rho_max}).
It is remarkable that the improvement of RSA with respect to SA is practically null for $d=20$ and very tiny for $d=100$.
While for $d=20$ one may claim that the improvement is absent because the model has a very weakly discontinuous phase transition (the range where the phase transition is continuous is very close by), for $d=100$ the 1RSB scenario holds clearly, but we do not see any improvement by reweighting states according to their internal entropy.
This observation raises some doubts about what RSA is actually doing and why is not working as expected.

Moreover, given that the performances of RSA are clearly worst than those of PT (see their algorithmic thresholds in Table~\ref{Tab:rho_max}), we arrive at the conclusion that there are more and less efficient ways to couple replicas.

\begin{figure}[t]
\centering
\includegraphics[width=0.47\textwidth]{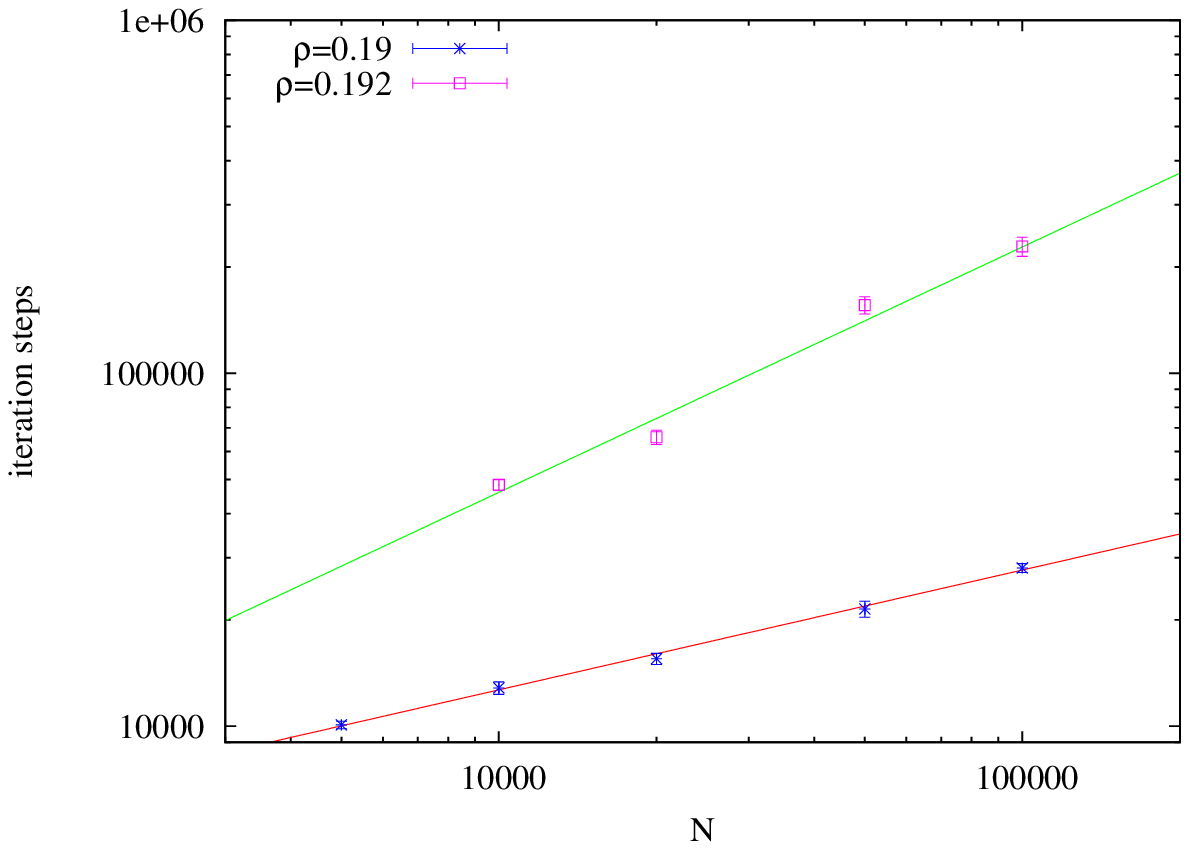}\hfill
\includegraphics[width=0.47\textwidth]{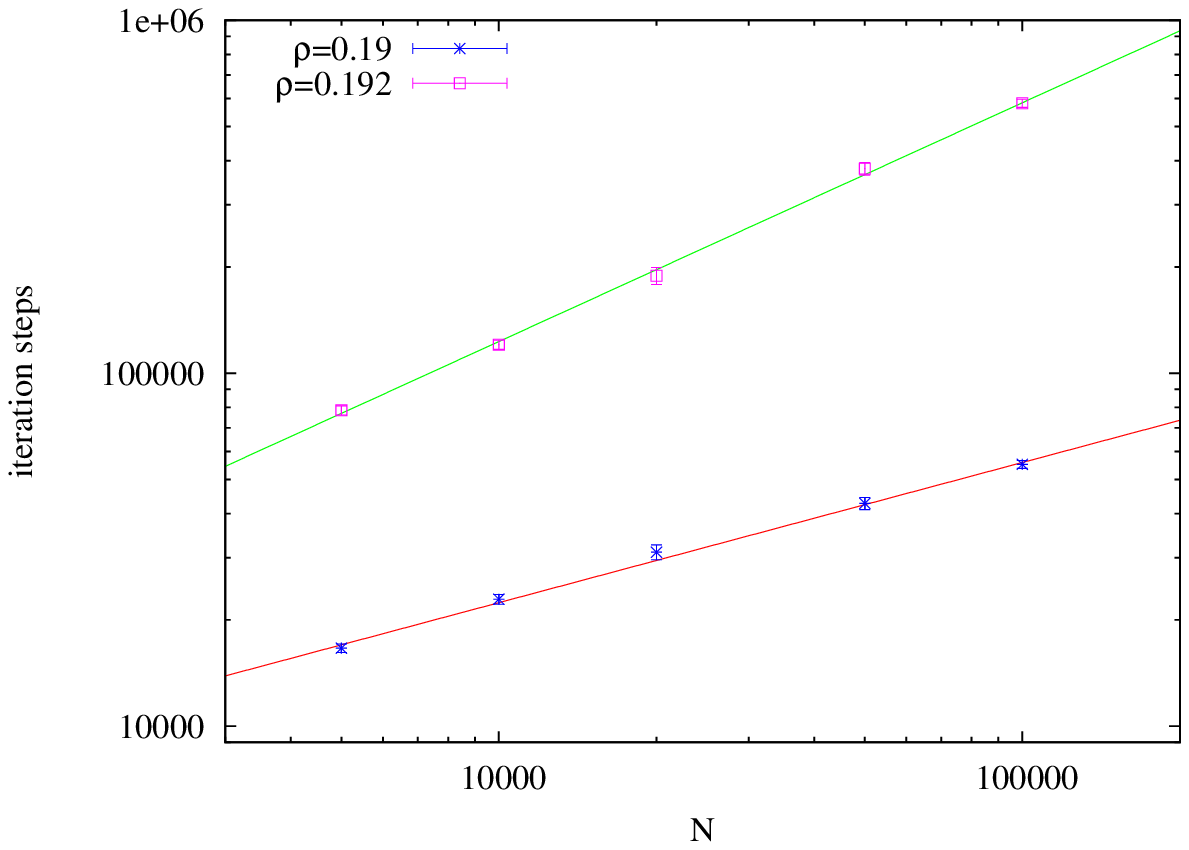}
\caption{\label{Fig:scalingNd20} Number of iterations to find a solution for $d=20$ and $\rho=0.19,0.192$ as a function of the problem size $N$ for the optimized $\beta$PT (left) and the $\mu$PT (right)
algorithms. The behaviour of the two algorithms is very similar.}
\end{figure}

\begin{table}[t]
\begin{center}\begin{tabular}{|l ||l|l||l| l |}
    \hline
 &$d=20$ $\rho=0.190$     &  $d=20$ $\rho=0.192$  &  $d=100$ $\rho=0.064$  &  $d=100$ $\rho=0.0646$  \\
     \hline
$\beta$PT & 0.339(8)  & 0.69(7) & 0.357(7)  & 0.42(2) \\
\hline
$\mu$PT & 0.40(1) & 0.68(1) & 0.336(8) & 0.44(3) \\
 \hline
 \end{tabular}
 \caption{The convergence time in Figs. \ref{Fig:scalingNd20} diverges as 
 $\tau(N)=a\cdot N^b$. In the table the comparison between values of $b$ for $\beta$PT and $\mu$PT}
 \label{Tab:b_comparison}
\end{center}
\end{table}

We move now to the analysis of the $\mu$PT algorithm. 
We use $N_\mu=21$ replicas evenly spaced by $\Delta\mu=0.2$ in the range $\mu\in[2,6]$ for $d=20$ and 
$N_\mu=31$ replicas evenly spaced by $\Delta\mu=0.15$ in the range $\mu\in[2,6.5]$ for $d=100$. 

We have already anticipated in Sec.~\ref{Sec:betaalg} that the behavior of $\mu$PT is equivalent to the one of $\beta$PT and in particular the algorithmic thresholds of the two algorithms are compatible. 
In Fig.~\ref{Fig:scalingNd20} we show that also the scaling of their running times with $N$ is similar, as the time needed to reach a solution of a given density scales as $\tau=a N^b$. In Table \ref{Tab:b_comparison} we make the comparison between the exponent $b$ of the two algorithms at the same values of $d$ and $\rho$.
These data confirm that the PT algorithm is a very robust one.

\section{Comparison with advanced Message Passing Algorithms} \label{Sec:BPR}

We have seen that Monte Carlo based algorithms easily outperform greedy algorithms and can reach densities well above the dynamical threshold $\rho_d$, passing also the rigidity threshold $\rho_r$ and for $d=20$ even beyond the condensation threshold $\rho_c$, thus approaching closely the maximum density $\rho_{max}$. This looks like a great result, but in order to put it under the right light, we need a comparison with a some other algorithm that is expected to work efficiently on this kind of optimization problems. 
Since the problem is defined on a random graph we expect message passing algorithms to be particularly well suited. For this reason, we have run also BPR on this problem.

\begin{figure}
\centering
\includegraphics[width=0.47\textwidth]{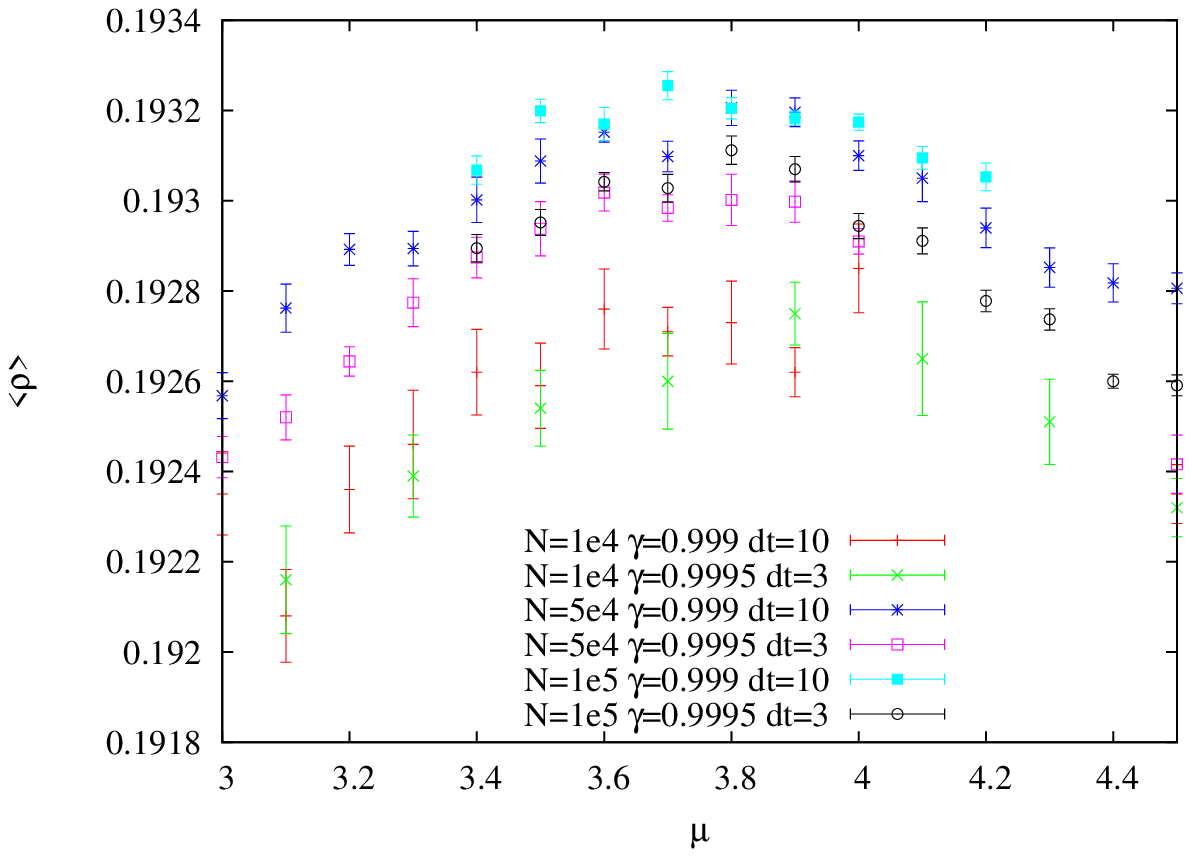}
\includegraphics[width=0.47\textwidth]{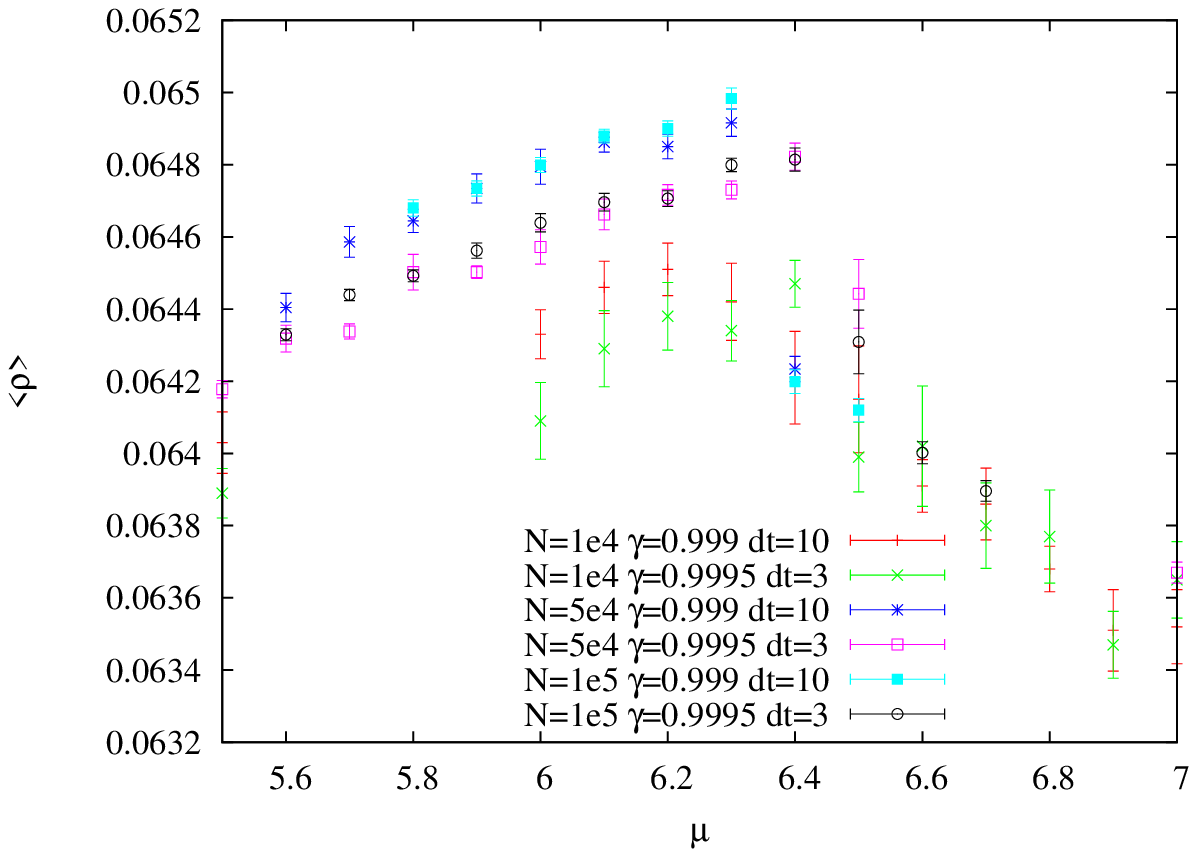}
\caption{\label{Fig:Reinf} Average density of the ISs found by BP+Reinforcement as a function of the chemical potential $\mu$ at different values of $N$ and of the BPR parameters.}
\end{figure}

In Fig.~\ref{Fig:Reinf} we show the average density of IS found by the BPR algorithm, as a function of the chemical potential $\mu$, for different values of the BPR parameters.
Let us just mention that below a certain chemical potential $\mu_L$, the solutions found by the BPR algorithm are always $n_i=0$, $\forall i$. 
The value of $\mu_L$ is the one that generates, using the RS solution of the model from \cite{barbier2013hard}, a density $\rho_L$ that roughly corresponds to the threshold density 
for the random vertex GA (for both $d=20$ and $d=100$ we have $\mu_L=2.15(5)$).

We have run the BPR algorithm in a broad range of chemical potentials and for different choices of the BPR parameters. 
The best results have been obtained with the choice $\gamma=0.999$ and $dt=10$. 
The maximum density reached can be deduced from the data shown in Fig.~\ref{Fig:Reinf} and it is clearly lower than the thresholds for the PT algorithms. 
Our best estimates are reported in Table~\ref{Tab:rho_max}. 
We notice that the threshold density for the BPR algorithm  is very similar to the one of the RSA algorithm and this is expected from Ref.~\cite{baldassi2016unreasonable}.

\section{Conclusions}

We have done a comparative study of algorithms to find the largest IS in a RRG of degree $d=20$ and $d=100$. Our aim was to understand the actual performances of different kind of algorithms (greedy, message passing and especially Monte Carlo based), and to connect their algorithmic thresholds with thermodynamical phase transitions. For both values of $d$ the set of IS undergoes a RFOT varying the IS density $\rho$, however for $d=20$ the transition is weakly discontinuous because of the vicinity to the range where the transition is continuous ($d<16$); for $d=100$ the transition is markedly discontinuous as in the large degree limit.

While Table~\ref{Tab:rho_max} summarizes thermodynamical and algorithmic thresholds, we list below the most relevant conclusions that we achieved:
\begin{itemize}
\item Only greedy algorithms get stuck below the dynamical threshold, while all the other algorithms easily pass beyond $\rho_d$; the relevance of the dynamical threshold for smart optimization algorithms seems very limited.
\item Also the condensation threshold at $\rho_c$ seems to play no role at all in describing the performances of the best optimization algorithms.
\item The simplest version of Monte Carlo algorithms seems to work roughly until the rigidity threshold at $\rho_r$, defined as the density where the time to diffuse away from a typical IS diverges.
\item More sophisticated Monte Carlo schemes (SA and PT) find IS beyond $\rho_r$, but without frozen variables, thus showing the ability of finding IS in atypical unfrozen states.
\item Replicated SA does not show any sensible improvement over standard SA for this problem, especially for $d=20$.
\item Belief Propagation with Reinforcement has an algorithmic threshold similar to Replicated SA.
\item Parallel Tempering is by far the best algorithm for solving this problem and can find IS of a very large density that no other algorithm can find.
\item Different versions of PT (in temperature and chemical potential) show almost the same algorithmic threshold, and this strongly suggests an universal behavior linked to an underlying phase transition. We conjecture the PT algorithmic threshold to coincide with the freezing threshold, i.e.\ PT is able to find an unfrozen IS as long as there is one.
\item Running times of PT are super-linear, but still polynomial in $N$. Algorithmic thresholds for super-linear algorithms are likely to be larger than those for linear algorithms, but a theory for the formers is completely lacking.
\end{itemize}

Our results clearly show the need for a theory for advanced Monte Carlo algorithms, like Parallel Tempering, which is at present lacking. Only by understanding analytically this class of algorithms we can hope to approach the ultimate algorithmic threshold for a broad class of hard optimization problems.

\begin{acknowledgments}
This research has been supported by the European Research Council under the European Unions Horizon 2020 research and
innovation programme (grant No.~694925 -- Lotglassy, G. Parisi).
\end{acknowledgments}

\bibliography{biblio}

\begin{thebibliography}{10}

\bibitem{bollobas1998random}
B{\'e}la Bollob{\'a}s.
\newblock Random graphs.
\newblock In {\em Modern graph theory}, pages 215--252. Springer, 1998.

\bibitem{hartmann2006phase}
Alexander~K Hartmann and Martin Weigt.
\newblock {\em Phase transitions in combinatorial optimization problems:
  basics, algorithms and statistical mechanics}.
\newblock John Wiley \& Sons, 2006.

\bibitem{bollobas1976cliques}
B{\'e}la Bollob{\'a}s and Paul Erd{\"o}s.
\newblock Cliques in random graphs.
\newblock In {\em Mathematical Proceedings of the Cambridge Philosophical
  Society}, volume~80, pages 419--427. Cambridge University Press, 1976.

\bibitem{frieze1990independence}
Alan~M Frieze.
\newblock On the independence number of random graphs.
\newblock {\em Discrete Mathematics}, 81(2):171--175, 1990.

\bibitem{coja2015independent}
Amin Coja-Oghlan and Charilaos Efthymiou.
\newblock On independent sets in random graphs.
\newblock {\em Random Structures \& Algorithms}, 47(3):436--486, 2015.

\bibitem{grimmett1975colouring}
Geoffrey~R Grimmett and Colin~JH McDiarmid.
\newblock On colouring random graphs.
\newblock In {\em Mathematical Proceedings of the Cambridge Philosophical
  Society}, volume~77, pages 313--324. Cambridge University Press, 1975.

\bibitem{gamarnik2014limits}
David Gamarnik and Madhu Sudan.
\newblock Limits of local algorithms over sparse random graphs.
\newblock In {\em Proceedings of the 5th conference on Innovations in
  theoretical computer science}, pages 369--376. ACM, 2014.

\bibitem{mezard2005clustering}
Marc M{\'e}zard, Thierry Mora, and Riccardo Zecchina.
\newblock Clustering of solutions in the random satisfiability problem.
\newblock {\em Physical Review Letters}, 94(19):197205, 2005.

\bibitem{achlioptas2011solution}
Dimitris Achlioptas, Amin Coja-Oghlan, and Federico Ricci-Tersenghi.
\newblock On the solution-space geometry of random constraint satisfaction
  problems.
\newblock {\em Random Structures \& Algorithms}, 38(3):251--268, 2011.

\bibitem{barbier2013hard}
Jean Barbier, Florent Krzakala, Lenka Zdeborov{\'a}, and Pan Zhang.
\newblock The hard-core model on random graphs revisited.
\newblock In {\em Journal of Physics: Conference Series}, volume 473, page
  012021. IOP Publishing, 2013.

\bibitem{achlioptas2006solution}
Dimitris Achlioptas and Federico Ricci-Tersenghi.
\newblock On the solution-space geometry of random constraint satisfaction
  problems.
\newblock In {\em Proceedings of the thirty-eighth annual ACM symposium on
  Theory of computing}, pages 130--139. ACM, 2006.

\bibitem{zdeborova2007phase}
Lenka Zdeborov{\'a} and Florent Krzaka{\l}a.
\newblock Phase transitions in the coloring of random graphs.
\newblock {\em Physical Review E}, 76(3):031131, 2007.

\bibitem{marino2016backtracking}
Raffaele Marino, Giorgio Parisi, and Federico Ricci-Tersenghi.
\newblock The backtracking survey propagation algorithm for solving random
  k-sat problems.
\newblock {\em Nature communications}, 7:12996, 2016.

\bibitem{braunstein2016large}
Alfredo Braunstein, Luca Dall’Asta, Guilhem Semerjian, and Lenka
  Zdeborov{\'a}.
\newblock The large deviations of the whitening process in random constraint
  satisfaction problems.
\newblock {\em Journal of Statistical Mechanics: Theory and Experiment},
  2016(5):053401, 2016.

\bibitem{feo1994greedy}
Thomas~A Feo, Mauricio~GC Resende, and Stuart~H Smith.
\newblock A greedy randomized adaptive search procedure for maximum independent
  set.
\newblock {\em Operations Research}, 42(5):860--878, 1994.

\bibitem{feo1995greedy}
Thomas~A Feo and Mauricio~GC Resende.
\newblock Greedy randomized adaptive search procedures.
\newblock {\em Journal of global optimization}, 6(2):109--133, 1995.

\bibitem{halldorsson1997greed}
Magn{\'u}s~M Halld{\'o}rsson and Jaikumar Radhakrishnan.
\newblock Greed is good: Approximating independent sets in sparse and
  bounded-degree graphs.
\newblock {\em Algorithmica}, 18(1):145--163, 1997.

\bibitem{zdeborova2010generalization}
Lenka Zdeborov{\'a} and Florent Krzakala.
\newblock Generalization of the cavity method for adiabatic evolution of gibbs
  states.
\newblock {\em Physical Review B}, 81(22):224205, 2010.

\bibitem{baldassi2016unreasonable}
Carlo Baldassi, Christian Borgs, Jennifer~T Chayes, Alessandro Ingrosso, Carlo
  Lucibello, Luca Saglietti, and Riccardo Zecchina.
\newblock Unreasonable effectiveness of learning neural networks: From
  accessible states and robust ensembles to basic algorithmic schemes.
\newblock {\em Proceedings of the National Academy of Sciences},
  113(48):E7655--E7662, 2016.

\bibitem{hukushima1996exchange}
Koji Hukushima and Koji Nemoto.
\newblock Exchange monte carlo method and application to spin glass
  simulations.
\newblock {\em Journal of the Physical Society of Japan}, 65(6):1604--1608,
  1996.

\bibitem{earl2005parallel}
David~J Earl and Michael~W Deem.
\newblock Parallel tempering: Theory, applications, and new perspectives.
\newblock {\em Physical Chemistry Chemical Physics}, 7(23):3910--3916, 2005.

\bibitem{moreno2003finding}
JJ~Moreno, Helmut~G Katzgraber, and Alexander~K Hartmann.
\newblock Finding low-temperature states with parallel tempering, simulated
  annealing and simple monte carlo.
\newblock {\em International Journal of Modern Physics C}, 14(03):285--302,
  2003.

\bibitem{angelini2018parallel}
Maria~Chiara Angelini.
\newblock Parallel tempering for the planted clique problem.
\newblock {\em Journal of Statistical Mechanics: Theory and Experiment},
  2018(7):073404, 2018.

\bibitem{dallasta2008entropy}
Luca Dall’Asta, Abolfazl Ramezanpour, and Riccardo Zecchina.
\newblock Entropy landscape and non-gibbs solutions in constraint satisfaction
  problems.
\newblock {\em Physical Review E}, 77(3):031118, 2008.

\bibitem{Karp1981Maximum}
R.~Karp and M.~Sipser.
\newblock Maximum matchings in sparse random graphs.
\newblock In {\em Proceedings of FOCS, Nashville, Tennessee, USA}, pages
  364--375, 1981.

\bibitem{wormald1995differential}
Nicholas~C Wormald et~al.
\newblock Differential equations for random processes and random graphs.
\newblock {\em The annals of applied probability}, 5(4):1217--1235, 1995.

\bibitem{braunstein2006learning}
Alfredo Braunstein and Riccardo Zecchina.
\newblock Learning by message passing in networks of discrete synapses.
\newblock {\em Physical review letters}, 96(3):030201, 2006.

\bibitem{aurell2004WALKSAT}
Erik Aurell, Uri Gordon, and Scott Kirkpatrick.
\newblock Comparing beliefs, surveys and random walks.
\newblock In {\em Proceedings of 17th NIPS}, page 804, 2004.

\bibitem{ardelius2006ASAT}
John Ardelius and Erik Aurell.
\newblock Behavior of heuristics on large and hard satisfiability problems.
\newblock {\em Physical Review E}, 74:037702, 2006.

\bibitem{Budzynski18}
Louise Budzynski, Federico Ricci-Tersenghi, and Guilhem Semerjian.
\newblock Biased landscapes for random constraint satisfaction problems.
\newblock {\em arXiv preprint:1811.01680}, 2018.

\bibitem{zdeborova2009statistical}
Lenka Zdeborov{\'a}.
\newblock Statistical physics of hard optimization problems.
\newblock {\em Acta Physica Slovaca. Reviews and Tutorials}, 59(3):169--303,
  2009.

\bibitem{montanari2003nature}
Andrea Montanari and Federico Ricci-Tersenghi.
\newblock On the nature of the low-temperature phase in discontinuous
  mean-field spin glasses.
\newblock {\em The European Physical Journal B-Condensed Matter and Complex
  Systems}, 33(3):339--346, 2003.

\bibitem{montanari2004cooling}
Andrea Montanari and Federico Ricci-Tersenghi.
\newblock Cooling-schedule dependence of the dynamics of mean-field glasses.
\newblock {\em Physical Review B}, 70(13):134406, 2004.

\bibitem{pelissetto2014large}
Andrea Pelissetto and Federico Ricci-Tersenghi.
\newblock Large deviations in monte carlo methods.
\newblock In {\em Large Deviations in Physics}, pages 161--191. Springer, 2014.

\end{thebibliography}

\appendix*
\section{Optimizing the choice of temperatures in the PT algorithms}\label{App}

Here we explain how we have implemented an optimized choice for the temperatures in the Parallel Tempering algorithm. We assume that the energy is close to its equilibrium value that can be computed via the RS solution.
For the $\beta$MC algorithm at a fixed density $\rho$, the RS mean energy can be computed noticing that the RS marginals are equal for each site and assume the value $p_{RS}(\sigma)=\rho\,\delta_{\sigma,1 }+(1-\rho)\delta_{\sigma,0}$, thus getting
\begin{equation}
e(\beta)=\frac{d}{2}\frac{\rho^2e^{-\beta}}{\rho^2e^{-\beta}+(1-\rho^2)}\;.
\end{equation}
In the large $N$ limit we can assume that the extensive energy at inverse temperature $\beta$ is a Gaussian variables with mean $E(\beta)=N e(\beta)$ and variance $\sigma^2(\beta)=-N e'(\beta)$. 
This Gaussianity assumption (which is rather well satisfied, but in the vicinity of the ground state) allows us to compute the probability of swapping two replicas at inverse temperatures $\beta_1$ and $\beta_2$,
\begin{equation}
p_\text{swap}(\beta_1,\beta_2) = \int dz_1\,dz_2\,\frac{e^{-z_1^2/2-z_2^2/2}}{2\pi} \min\left(1,e^{(\beta_2-\beta_1)(E(\beta_2)+\sigma(\beta_2)z_2-E(\beta_1)-\sigma(\beta_1)z_1)}\right)\;.
\end{equation}
In the limit $\Delta\beta=\beta_2-\beta_1\ll 1$ we can approximate $E(\beta_2)-E(\beta_1)\simeq N e'(\beta) \Delta\beta$ with $\beta=(\beta_1+\beta_2)/2$ and $\sigma(\beta_1)\simeq\sigma(\beta_2)\simeq\sigma(\beta)=\sqrt{-Ne'(\beta)}$, thus getting
\begin{equation}
p_\text{swap}(\beta,\Delta\beta) = \int dz_1\,dz_2\,\frac{e^{-z_1^2/2-z_2^2/2}}{2\pi} \min\left(1,e^{\Delta\beta[N e'(\beta) \Delta\beta+\sqrt{-Ne'(\beta)}(z_2-z_1)]}\right) =\text{erfc}\left(\frac{\Delta\beta\sqrt{-Ne'(\beta)}}{2}\right)\;.
\end{equation}
The best way to allow replicas to wander fast between temperatures is to fix a constant $p_\text{swap}$ between any pair of successive temperatures and this can be achieved with the choice
\begin{equation}
\beta_{n+1} = \beta_n + \frac{r}{\sqrt{N |e'(\beta_n)|}}
\label{eq:betasPT}
\end{equation}
implying $p_\text{swap}=\text{erfc}(r/2)$. The optimal value for $r$ can be obtained by maximizing the mean squared distance traveled by a random walker performing jumps of size $r$ with probability $\text{erfc}(r/2)$, that is
\begin{equation}
r_\text{opt} = \text{argmax} \left[\text{erfc}\left(\frac{r}{2}\right) r^2\right] \simeq 1.68376\;,
\end{equation}
leading to an optimal swapping rate equal to $\text{erfc}(r_\text{opt}/2) \simeq 0.23381$ (this is the well-known 0.23 rule \cite{pelissetto2014large}).

With the set of temperatures defined in Eq.~(\ref{eq:betasPT}) the optimized PT would require $O(\sqrt{N})$ replicas. However, we empirically observe that the time of convergence of the algorithm does not change if replicas in the range $\beta\in[0,\beta_{min}]$ are removed. We find empirically that the largest possible value for $\beta_{min}$ roughly corresponds to the inverse temperature at which the equilibrium magnetization coincide with the maximum IS density reached by the greedy algorithm, $\rho_{GA}$. This is very reasonable, indeed for $\rho<\rho_{GA}$ we do not expect any relevant barrier to be present and so the PT replicas at $\beta_{min}$ can easily travel the entire configurational space.

For $\beta_{max}$ we choose the lowest inverse temperature at which the condition $E(\beta)-\sqrt{N |e'(\beta)|}<0$ is satisfied, implying that a typical spontaneous fluctuation can lead the algorithm to find a configuration of zero energy.

We observe that the optimized version of $\beta$PT finds solutions up to a $\rho_{max}(d=20)=0.1941(5)$, compatible with the non-optimized version, but with a smaller exponent $b=2.4(3)$. For $d=100$ the optimized $\beta$PT algorithm reaches $\rho_{max}(d=100)=0.06572(9)$ with $b=3.2(1)$.

Things are different for the $\mu$PT algorithm. For this algorithm, the RS magnetization is the one written in Eq.~(6) of Ref.~\cite{barbier2013hard}. However, in the hard region, the real magnetization it is quite different and so we can not use the RS result to optimize the $\mu$PT algorithm.

\end{document}